# Low-temperature environments for quantum computation and quantum simulation

Hailong Fu[1*], Pengjie Wang[2*], Zhenhai Hu[3], Yifan Li[3], Xi Lin[3,4,5]

1 Department of Physics, The Pennsylvania State University, University Park, Pennsylvania 16802, USA
2 Department of Physics, Princeton University, Princeton, New Jersey 08544, USA
3 International Center for Quantum Materials, Peking University, Beijing 100871, China
4 Beijing Academy of Quantum Information Sciences, Beijing 100193, China
5 CAS Center for Excellence in Topological Quantum Computation, University of Chinese Academy of Sciences, Beijing 100190, China
*These two authors contributed equally to this work.

*Abstract: This review article summarizes the requirement of low temperature conditions in existing experimental approaches to quantum computation and quantum simulation.*

**1 introduction**

Since the possibility of quantum computation (QC) was proposed around forty years ago [1-5], this concept has been appealing and expected to solve problems insoluble for classical computers, whose calculation capacity has been rocketing but is clearly anticipated to reach its limit when the size of transistors is approaching the size of atoms. In a quantum computer, a quantum bit (qubit) replaces the bit in a classical computer. And quantum logic gates, unitary transformations acting on the qubits' Hilbert space, take the place of classical logic gates [6, 7]. As an example of controllable quantum systems, qubits can also be used to simulate other quantum systems, and such an approach to simulating quantum systems with a more controllable quantum system is called quantum simulation (QS) [2, 8, 9].

In principle, the basic implementation of QC can utilize any particular design of qubits, behind which there are two vector states as bases in the Hilbert space. We would prefer such a qubit to be convenient to be initialized, read out and corrected, and more importantly, to be easy to scale up. However, there is no agreement on which physical system is the best candidate for an ultimate qubit. Enormous efforts have been made on the realization of QC from quite diverse approaches [7, 10-17], and one would expect that different approaches may be important at different stages in the development of QC. As a comparison, in the improvement of data storage in classical computers, the dominant technique has been switching among floppy disk, SSD (solid-state drive), CD (compact disc), Zip drive, DVD (digital versatile disc), and USB (universal serial bus) flash drive, and each of them was important and popular in a particular period of time.

Notably, a typical condition in some experimental schemes of QC and QS is the requirement for low temperature environments, which have revealed a wide array of fascinating quantum phenomena, and lots of them were unexpected before their discoveries, such as superfluidity [18], superconductivity [19], quantum Hall effect [20], and fractional quantum Hall effect [21]. Therefore, it would be interesting to go through the environmental temperature requirements for different realistic implementations of qubits: although the operation at room temperature is admirable, the requirement of a low temperature cryostat is also affordable.

## 2 simplified picture of cryogenics

The environmental temperature in this review refers to the cryostat temperature, which is provided by a thermal bath that not only cools itself down but also maintains the capacity to cool other macroscopic objects for general purposes. There are many physical principles for cooling, including the well-known evaporation and gas expansion. An environmental temperature $\geq 77$ K can be achieved with the help of liquid nitrogen. In order to reach 4 K, liquid $^4$He, the major isotope of helium, was usually needed. Compared with liquid nitrogen, liquid $^4$He is of lower heat capacity, smaller latent heat of evaporation, and higher price. Liquid $^4$He comes from the liquefaction of $^4$He gas, which is usually purified from the byproduct of natural gas [22], or from the recycled $^4$He gas. The supply of $^4$He is limited for many different reasons, and there have been several shortages since 2006 [23], which impacted the scientific community severely.

Because of the technological advance of Gifford–McMahon [24] and pulse-tube [25] cryocoolers, 4 K environments can be easily accessed and maintained without the need of continuous supply of liquid $^4$He or formal training in cryogenics for operators. Cryocoolers can provide a cooling power of ~1 W (typically 0.1-2 W for commercial cryocoolers) at 4.2 K, which roughly equals to liquid $^4$He consumption of 30 liters per day. New crisis of helium shortage in a few years is still expected, so the cryocoolers based on electricity are apparently appealing. In addition, although demands for low temperature environments from the fundamental research and promising applications of QC and QS are quickly increasing, the field of helium physics, which trains part of low temperature specialists conventionally, has been shrinking for decades. Therefore, cryocoolers, which can continuously operate without cryogen transfer and specialized training in cryogenics, are of unbeatable advantage.

The historic point in the development of cryocoolers was the time reaching 4.2 K [26], above which they can replace the role of liquid $^4$He cryogen in principle. The refrigeration to lower temperatures is usually based on a ~ 4 K precooling environment, in which cryocoolers can replace liquid $^4$He easily, with a minor modification of existing designs. Evaporation of $^4$He or $^3$He, phase separation of $^3$He-$^4$He mixture, adiabatic demagnetization, and solidification of liquid $^3$He can all be used to approach lower temperatures from a 4 K precooling environment [27]. Although those methods may require $^3$He and $^4$He, helium is treated as a non-consuming refrigerant, while liquid $^4$He for 4 K precooling is used as disposable "fuel" of the cooling reservoir. The refrigerators based on cryocoolers are more suitable for the applications of QC and QS, for the reason of non-interrupted and more user-friendly operations. Gifford–McMahon cryocoolers are of higher efficiency and lower price. However, there are less vibrations and requirements for maintenance in pulse-tube cryocoolers. Therefore, the pulse-tube type is more popular in ultra-low temperature cryostats. Typically, the ultra-low temperature refers to temperatures below 1 K or 0.3 K.

## 3 experimental approaches

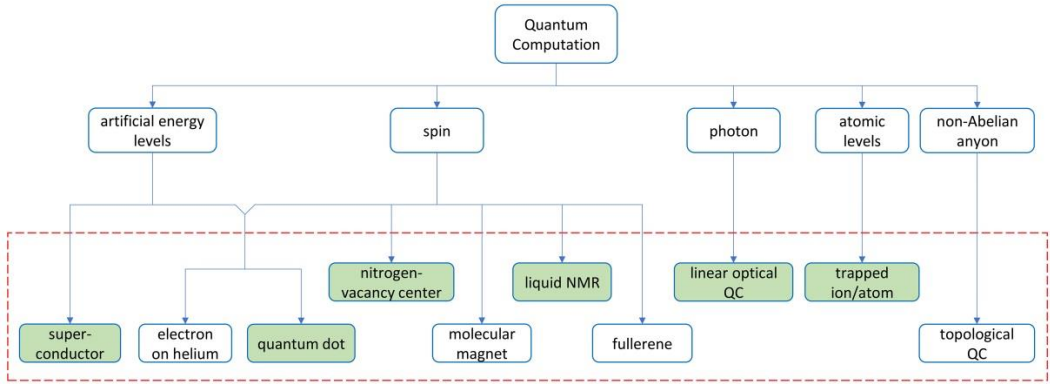

**Figure 1:** A rough classification of QC based on the physical principle of the qubit. The level below the box "Quantum Computation" indicates the major degree of freedom which is utilized to store and process quantum information. The terms circled by the red dashed box are examples of practical systems. Nuclear magnetic resonance is shortened as NMR. QS experiments have been reported in systems with green color background.

In Fig.1, we try to classify some of the existing quantum computing systems based on the degree of freedom of qubits: apparently, the combination of different degrees of freedom in an experimental scheme is common. The essential component of a qubit is the existence of superposition states, which can originate from natural energy levels, artificial energy levels, and degenerate non-Abelian states. Natural energy levels can rely on atomic levels, or properties of spin and photon. Qubits utilizing atomic levels can be realized in trapped ion/atom systems. The polarization of photons, the number of photons, or spins of electrons and nuclei are also convenient to define different quantum states. Secondly, energy levels can be created artificially. The charge, phase, and flux superconducting qubits all share the same characteristics of artificial energy potential levels. Quantum dots are famous as artificial atoms, and the system of electrons on helium is another example with artificial levels. It should be noted that electrons in quantum dots or on top of helium also share the degree of freedom arising from the natural property of spin. Lastly, the braiding of non-Abelian anyons with degenerate ground states can also be used for QC, which is special for its potential fault-tolerant application.

Well-known practical systems for QC are provided in the red dashed box of Fig. 1. The environmental temperature ranges of different systems are summarized in Fig. 2. In the following sub-subsections, we will briefly introduce how to achieve the needed low temperature environments.

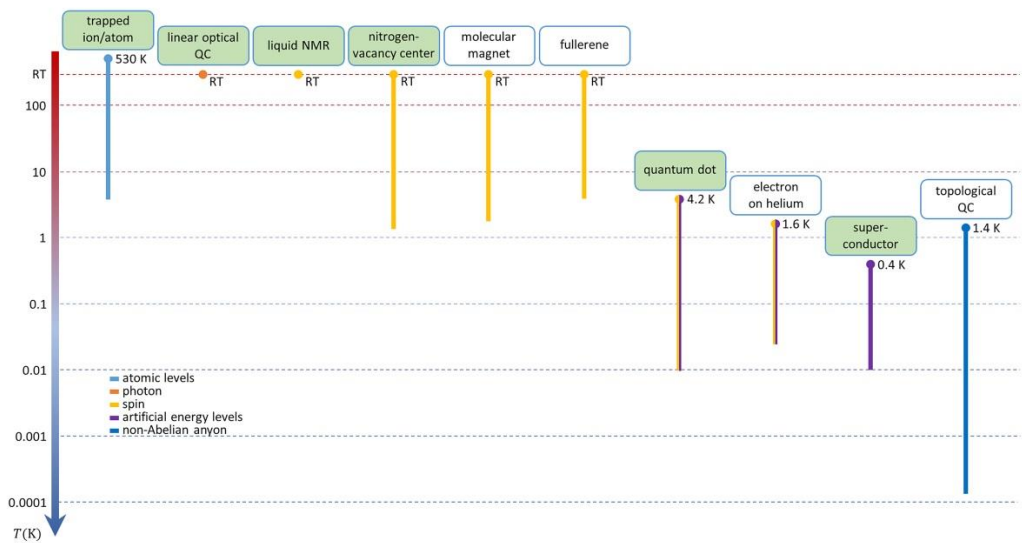

**Figure 2:** The possible operating temperature ranges of physical QC systems discussed in this work. RT stands for room temperature. The temperature here refers to the environmental temperature. The vertical strips indicate the temperature ranges of the corresponding systems, and the highest related temperature we are aware of for each system is marked with a dot. The colors of strips/dots indicate the major degree of freedom, as shown in the legend. The corresponding temperature limits are from references: trapped ion/atom (upper: [28], lower: [29]), linear optical QC [30], liquid NMR [31], nitrogen-vacancy center (upper: [32, 33], lower: [34]), molecular magnet (upper: [35], lower: [36]), fullerene (upper and lower: [37]), quantum dot (upper: [38], lower: [39, 40]), electron on helium (upper: [41], lower: [42]), superconductor (upper: [43], lower: [44]), and topological QC (upper: [45], lower: [46]). QS experiments have been reported in systems with green color background.

## 3.1 liquid nuclear magnetic resonance, nitrogen-vacancy center, trapped ion/atom, and linear optical QC

The liquid nuclear magnetic resonance method [10, 31, 47-65] uses the qubits based on nuclear spins in the solution of molecules, and probes the quantum states by nuclear magnetic resonance spectroscopy. A nitrogen-vacancy center in diamond consists of a nitrogen impurity and a carbon vacancy. Localized around the vacancy, a spin-triplet system is formed with two unpaired electrons, generating a subspace where a qubit can reside [11, 32-34, 66-81]. A qubit can also be constructed using the atomic levels of trapped ultra-cold particles, among which the most well-studied examples are ions trapped by electromagnetic fields [12, 82-85] and neutral atoms in an optical trap [86-90]. Optical quantum systems are prominent candidates for QC, which use photons as qubits. The scheme of QC based on optical systems initially involved nonlinear optics, such as cavity quantum electrodynamics [91, 92], while the linear optical method [93] was reported in 2001.

Although the magnetic field for liquid nuclear magnetic resonance may be generated by a superconducting solenoid, which requires a low temperature environment, the liquid sample itself is operated at room temperature. In a nitrogen-vacancy center, the low expansion coefficient in diamond makes coherence time weakly dependent on temperature [94] and reach up to milliseconds (seconds) for electron (nuclear) spin at room temperature [95, 96]. Similarly, all the operations in the linear optical QC can basically function at room temperature.

There is some subtlety in the concept of temperature for trapped ions/atoms. The trapped particles are often cooled to an effective temperature of ~ μK, but this effective temperature is defined according to the average kinetic energy of the particles. The trapped particles, however, are not necessarily in thermal equilibrium, and thus may not have a well-defined temperature. The environmental temperature addressed in this review refers to a well-defined temperature provided by a thermal bath, which is capable of cooling other macroscopic objects. In this sense, for trapped ion/atom systems, we treat the temperature of the trap, instead of the effective temperature of the trapped particles, as the environmental temperature. Trapped ion/atom systems can function properly with the trap at room temperature or even higher [12, 28]. However, low-temperature traps are still advantageous for their lower noise and better performance [29, 97-101].

## 3.2 molecular magnet and fullerene QC

The spin qubits could be stored in the molecular magnet system [102-104] and the fullerene system [13, 37, 105]. The molecular magnet system is known as molecular clusters with giant anisotropic spins or a molecule with a single magnetic ion, while the fullerene system could cage a single atom (for example, $^{14}N$, $^{15}N$, $^{31}P$) with atomic spin. The studies of the molecular magnet system and the fullerene system are still within fundamental research. In the molecular magnet system, a coherence time higher than 1 μs has been observed in several materials from 2 K to room temperature [35, 36, 106-111], while in the fullerene system, a coherence time of 139 ns has been reached at 200 K [37].

To achieve low enough operation temperatures for molecular magnet and fullerene QC, the following methods may be adopted. Liquid nitrogen and liquid $^4$He can be used as cryogens to provide ~ 77 K and ~ 4 K environments conveniently. Combination of cryogens, vacuum, and heating with negative feedback can efficiently maintain an environment from 300 K to 4 K. Flow refrigerators through which liquid nitrogen or liquid $^4$He flows with adjustable rates can also maintain an environment higher than 77 K or 4 K, although flow refrigerators consume more cryogens. One can also achieve environmental temperatures above 4 K using cryocoolers without cryogen consumption. The cooling power of commercial pulse-tube cryocoolers, which are more popular for ultra-low temperature refrigerators, increased from ~ 0.5 W to ~ 2 W at 4.2 K in roughly 15 years. Currently, commercial cryocoolers can stably reach temperatures below 3 K.

### 3.3 quantum dot, electron on helium, and superconducting QC

Charge and spin qubits can be formed in quantum dots where discrete quantum levels are resolvable, and one electron can be trapped into the lowest available level at low temperatures [112, 113]. Charge qubits are sensitive to the surrounding charge noise [114-116], while spin qubits could be less coupled to the environment and generally have longer coherence time than charge qubits do [117-119]. The study of quantum dot qubits was first initiated in GaAs/AlGaAs heterostructures [120-123], where the two-dimensional electron gas is of high quality. However, the hyperfine interaction between the electron spin and the nuclear spin severely suppresses the coherence time of spin qubits, which limits the improvement of the fidelity [122, 124-128]. Therefore, IV semiconductors with weak hyperfine interaction are promising candidates for spin qubits, and a lot of progress has been made in silicon-based [39, 129-134] and germanium-based [135-140] spin qubits. In silicon quantum dot spin qubits, a dephasing time of $T_2^* = 120$ μs has been recorded in an isotopically purified $^{28}$Si CMOS-like device [129]. Beside the electron spin in silicon quantum dots, the electron and nuclear spins of phosphorus ($^{31}$P) donors in silicon can also act as qubits with very long coherence time [141-144]. Valley-spin and charge qubits have been realized in carbon nanotube [145-148], and the study of graphene quantum dot qubits is also in progress [149-152]. High flexibility in device fabrication, the advantage of combination with other two-dimensional materials, and possible artificial synthesis with pure $^{12}$C could make graphene-based qubits competitive in the future [149, 153]. The study of quantum dot qubits is also progressing in other systems such as nanowires [154-156], van der Waals transition metal dichalcogenides [157-159], and self-assembled quantum dots [38, 160, 161]. Benefited by the modern semiconductor fabrication technology for classical computers, quantum dot qubits are considered to be highly scalable and promising for QC in the future [162-164], though the qubit number is still developing [40, 165, 166].

The qubits based on electrons on helium [167, 168] are very similar to those based on quantum dots. Both the two lowest energy levels and the spin of the electron can be utilized to construct qubits [167-171]. The electrons can stay on the surface of superfluid $^4$He, which may hold the record of the highest mobility among two-dimensional transport systems [172].

Superconducting qubits are formed in superconducting quantum circuits comprising capacitors, inductors, and Josephson junctions, where artificial energy levels can be engineered [15, 173, 174]. With different relationships between Coulomb energy and Josephson energy, superconducting quantum circuits can host qubits with charge, flux or phase degree of freedom [175, 176]. The first experimental demonstration of a superconducting qubit, a Cooper-pair Box (a charge qubit) with nanosecond-scale coherence time was realized in 1999 [177]. Flux qubit and phase qubit came along in the next few years [178, 179]. The transmon qubit [180-185], developed from the Cooper-pair Box, usually has longer coherence time because of effectively reduced charge noise sensitivity and is a popular qubit design used in the scaling up of QC today. Recently, the coherence time in a 3D transmon qubit has reached around 200 μs [185]. The exploration of different kinds of qubits is still ongoing [15, 44, 186-197]. The number of qubits has been increasing continuously in the last decade [198-213] as shown in Fig. 3. In 2019, Google reported the demonstration of quantum supremacy using a 53-qubit processor (named Sycamore) and claimed it could solve a problem dramatically faster than a classical computer [209, 214]. In the meantime, progress on benchmarking multi-qubit devices has also been made [202, 206, 207, 215-217]. Recently, 18-qubit genuine multipartite entangled states in 20-qubit devices have been demonstrated [206, 207]. As for the superconducting materials, aluminum is generally used in superconducting QC. Compared with niobium, aluminum with lower melting point makes qubit fabrication process simpler and more flexible, and high-quality aluminum/amorphous-aluminum-oxide/aluminum Josephson junctions guarantee longer coherence time [218-221]. As a result, aluminum-based qubits are preferred for building large-scale quantum computers at present, even though niobium has a higher superconducting transition temperature.

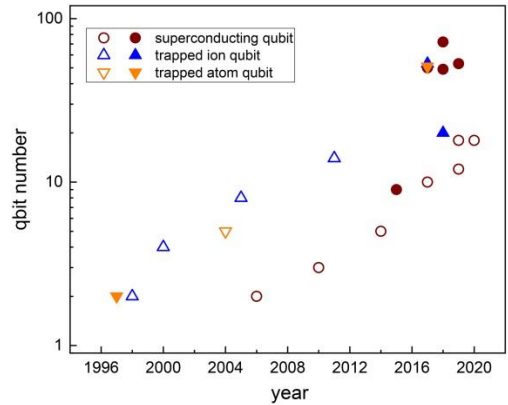

**Figure 3:** Evolution of qubit number in superconducting, trapped ion, and trapped atom systems. Open symbols represent the number of qubits used in preparing a specific state with demonstrated genuine multipartite entanglement (fidelity > 0.5). Solid symbols represent the number of qubits without verified all-qubit genuine multipartite entanglement. Data are from references: trapped ion qubit [222-227], trapped atom qubit [228-230], and superconducting qubit [198, 200, 202-204, 206, 207, 209-213].

Fault-tolerant quantum computers [231] are possible if gate fidelities are higher than certain thresholds [232-236], typically ranging from ~ 99% [237, 238] to ~ 99.9999% [234-236, 239] for different models. High fidelity quantum gates have been demonstrated in trapped ions, with a single-qubit gate fidelity up to 99.9999% [240] and a two-qubit gate fidelity larger than 99.9% [241, 242], approaching or even surpassing the fault-tolerant thresholds. Recently, a mixed-species two-qubit gate with a fidelity of 99.8% is realized [243], comparable to the best fidelity achieved with single ion species. Trapped atoms show lower yet competitive gate fidelities: 97% for two-qubit gates [244], and ≥ 99.99% for single-qubit gates [245]. Single-qubit gates with a fidelity of > 99.99% and CNOT gates with a fidelity of 99.2% have been realized in nitrogen-vacancy center [33]. In silicon quantum dot spin qubits, single-qubit gates with a fidelity of ≥ 99.9% [131] and two-qubit gates with a fidelity of ~ 98% [132] have been achieved. Recently, the operation in silicon quantum dot qubits above 1 K has been reported [133, 134], which could facilitate the realization of large scale silicon-based quantum computers. Qubits based on phosphorus ($^{31}$P) donors in silicon have single-qubit gate fidelities up to 99.95% (electron spin) and 99.99% (nuclear spin) [143]. Single-qubit gates with a fidelity of 99.3% and two-qubit gates based on hole spin qubit in germanium have also been achieved recently [138]. In superconducting QC, a two-qubit gate was reported in 2006 [198], as early as we are aware of, and fidelities of ≥ 99.5% have been realized as of today [246, 247]. Despite the impressive progress on different kinds of high-fidelity operations, the realization of truly fault-tolerant QC still needs quantum error correction [248], which usually encodes logical qubits using multiple physical qubits. To date quantum error correction is still at an early stage [44, 50, 203, 249-255], and the scheme of topological QC also seems to be attractive to some researchers.

The research on quantum dot, electron on helium, and superconducting QC usually requires ultra-low temperature environments. Evaporation of liquid $^3$He is a convenient method to reach the ultra-low temperature regime. Pre-liquefying the gas $^3$He, accumulating the liquid $^3$He, and then evaporating the liquid $^3$He can provide an environment of ~ 300 mK in practice, although most of the $^3$He refrigerators' base temperatures are higher than 300 mK. Once the $^3$He atoms are removed from liquid $^3$He through liquid $^4$He rather than through vacuum, where the phase separation of $^3$He-$^4$He mixture is required, the minimum temperature can be lower than 10 mK, and such cryostats are named as dilution refrigerators [27, 256]. Dilution refrigerators are the mainstream refrigerators for QC and QS. For example, commercial companies such as IBM, Google, Intel, D-Wave, Rigetti, etc., have been announcing QC hardware operated in dilution refrigerators. Qubit devices are thermally anchored on the mixing chamber plate (the position with the lowest temperature, where $^3$He-$^4$He phase separation occurs) of dilution refrigerators and connected to electronics at room temperature through control and readout wires. Noise from the instruments and the environment can severely reduce the qubit coherence time and the fidelities of operations [7, 257, 258]. Hence the filtering system in dilution refrigerators is essential for qubit operation. Usually low-pass filtering is installed in direct current wires, and microwave components such as attenuators, filters and amplifiers are installed in radio-frequency wires [259, 260].

Making use of the special minimum in the melting curve of $^3$He, the solidification of $^3$He can be a cooling source under appropriate conditions, named as Pomeranchuk cooling [261, 262]. Pomeranchuk cooling can work to as low as around 2 mK, famous in history for hosting the superfluid $^3$He experiments [263, 264]. Adiabatic demagnetization of paramagnetic salts can also provide an environment below 100 mK, where an external field is used to determine the order of electronic magnetic moments [265]. The entropy of magnetic moments can be tuned by the external magnetic field, and cooling happens with the following steps: an isothermal magnetization with pre-cooling, a thermal isolation from the pre-cooling environment, and an adiabatic demagnetization. Adiabatic demagnetization with paramagnetic salts can cool to temperatures almost as low as dilution refrigerators can do. However, Pomeranchuk cooling and adiabatic demagnetization are "one-shot" methods that cannot provide persistent low temperature environment, so they are inappropriate to be directly used for supporting QC and QS.

**3.4 topological QC**

Topological QC was proposed by Kitaev in 2002 to solve the problems of decoherence and error correction in QC [266]. Topological QC exploits non-Abelian anyons to store and manipulate quantum information through braiding, which represents the exchange of identical anyons in low dimensions. The braiding of anyons will not be perturbed by small local noise from the environment due to its topological properties, which makes topological QC unique in theory. In practice, however, error can still be introduced into the system [17, 267-269], such as the thermal fluctuations producing random stray pairs of anyons [267, 269].

To identify the systems hosting non-Abelian anyons, many attempts have been made in the past few decades. As the first proposed system, the fractional quantum Hall state at filling factor 5/2 may be a Pfaffian state with non-Abelian quasiparticle excitations [267, 270-275]. Other wave functions for the non-Abelian 5/2 state have also been proposed but there are only limited experimental evidences to distinguish among them [273-280]. Majorana zero modes, which may be located at the boundaries and defects in topological superconductors, share the universal properties with the Pfaffian state, and are also predicted to support non-Abelian statistics. The practical systems include spin-orbit couple involving superconductor [17, 281-300], topological insulator coupled with superconductor [301-313], and more recently, vortex bound state of iron-based superconductors [314, 315], etc. However, anyons obeying non-Abelian statistics are not necessarily capable of universal quantum computation [16]. As a possible solution, Fibonacci anyons and other non-Abelian anyons are proposed [16, 269, 316-323]. Table 1 provides some examples of possible experimental platforms for topological QC.

Although the topological systems are promising for the fault-tolerant QC, the braiding itself is challenging. The research of topological QC still remains in the field of fundamental research.

Nevertheless, the realization of practical QC is still obscure at this moment. It is difficult or even impossible to conclude which direction of QC research is unworthy. In such sense, the study of topological QC with the long-term benefit of fault-tolerant property, though still at an early stage, should be encouraged.

Usually the studies of topological QC are performed with dilution refrigerators, but lower temperature is beneficial for its exploration. Adiabatic nuclear demagnetization, usually pre-cooled by a dilution refrigerator, is the method to reach the lowest temperature in the concept of condensed matter, and it is the only method to cool macroscopic objects to microkelvin regime. Different from adiabatic demagnetization of paramagnetic salts, nuclear magnetic moments are used in adiabatic nuclear demagnetization rather than electron magnetic moments [324]. The environmental temperature provided by an adiabatic nuclear demagnetization refrigerator can be below 20 μK [325]. Studies related to the Majorana modes and Andreev bound states in superfluid $^3$He have been carried out with adiabatic nuclear demagnetization for the lower temperature condition [46, 326-332].

| Possible systems for topological QC | | | Examples of materials |
|---|---|---|---|
| **FQH related** | **SC related** | **TI related** | |
| even-denominator FQH state | | | GaAs/AlGaAs [267, 270, 272-275, 333] |
| 12/5 FQH state | | | GaAs/AlGaAs [334] |
| confined FQH edge | | | GaAs/AlGaAs [335, 336] |
| FQH edges coupled to SC | | | GaAs/AlGaAs + SC [316, 322, 337-339] |
| | intrinsic topological SC | | Sr$_2$RuO$_4$ [340, 341]<br>Cu$_x$Bi$_2$Se$_3$ [342, 343]<br>Cu$_x$(PbSe)$_5$(Bi$_2$Se$_3$)$_6$ [344] |
| | spin-orbit couple involving SC | | (InAs or InSb) + SC [292-296]<br>EuS/Au + SC [299] |
| | magnetic surface adatoms or underlying islands with SC | | Fe atomic chain on Pb [345-347]<br>Fe or Co atomic island on Pb [348, 349]<br>Fe on Re(0001)-O(2×1) [350] |
| | 2DEG coupled with SC | | Proposal only [351] |
| | topological insulator coupled with SC | | Bi$_2$Te$_3$ + SC [303-305]<br>HgTe related + SC [306-308]<br>InAs/GaSb Quantum Well + SC [309-311]<br>Fe/Bi + Nb [312] |
| | vortex bound state of iron-based SC | | Fe(Te, Se) [314, 315] |
| | quantum anomalous Hall insulator coupled with SC | | (Cr$_{0.12}$Bi$_{0.26}$Sb$_{0.62}$)$_2$Te$_3$ + SC [352, 353] |
| | | Non-Abelian Jackiw-Rebbi(-like) modes in TI | Proposal only [354, 355] |
| **Others** | | | |
| quantum spin liquid | | | α-RuCl$_3$ [356-358] |
| superfluid $^3$He | | | $^3$He [46, 329-332, 359] |
| interacting nanowires with/without SC | | | Proposal only [360-362] |
| Majorana Kramers pair in time-reversal invariant topological SC | | | Proposal only [363, 364] |

**Table 1:** Some possible systems for topological QC. FQH refers to fractional quantum Hall, SC refers to superconductor, and TI refers to topological insulator.

## 3.5 QS

Quantum simulators [8] are quantum processors performing QS and can be loosely divided into three categories: analog, digital, and hybrid. An analog quantum simulator emulates a quantum system by mapping the Hamiltonian of the system onto the Hamiltonian of the quantum simulator. The experimental implementation of analog quantum simulators is typically based on qubits, though in principle this is not necessary. In a digital quantum simulator, the dynamics or static properties of a quantum system are reconstructed with quantum logic gates operating on qubits. Such a simulation can probably be classified as an application of QC as well. Moreover, hybrid digital-analog quantum simulators are also possible [365-369]. In the following, we will briefly review QS realized with different categories of quantum simulators. Further reading about QS can be found in references [8, 9, 368, 370-377].

Since the early stages of QS, the analog quantum simulators have been widely studied for a large diversity of quantum systems from condensed matter physics to high-energy physics, with various physical implementations including liquid nuclear magnetic resonance systems [378, 379], linear optical systems [380], trapped ions [226, 381-393] or atoms [230, 394-397], superconducting qubits [398-403], quantum dots [404, 405], and nitrogen-vacancy centers [406]. Although analog quantum simulators are less sensitive to decoherence and easier to scale up, they are problem-specific, contrary to digital quantum simulators which can universally simulate any finite local Hamiltonians in principle [407]. Early prototypes of digital quantum simulators were realized with liquid nuclear magnetic resonance systems [408-414] due to the well-developed control techniques. Later, digital quantum simulators have also been realized with linear optical systems [415, 416], trapped ions [417-422], nitrogen-vacancy centers [423], and superconducting qubits [424-430]. Digital quantum simulation has been performed in a superconductor-based quantum processor with 20 qubits developed by IBM [429]. More recently, on Google's Sycamore processor [209], a quantum chemistry simulation using up to 12 qubits has been achieved [430]. Hybrid digital-analog quantum simulators have also been proposed [365-368], with the possibility of combining the advantages of both digital and analog quantum simulators. Experiments for hybrid simulators have been done with superconducting qubits [369]. Among the various physical implementations of QS, superconducting qubits and quantum dots are operated in dilution refrigerators, in need of low-temperature environments.

Commercial companies have also been active in the field of QC and QS, and QS has been demonstrated in some prototypes of quantum processors invented by commercial companies. D-Wave company has invented and commercialized several generations of annealing-based quantum processors, and announced a processor with more than 5000 qubits recently [431]. The D-Wave processors perform quantum annealing [432], a quantum algorithm that analogically optimize a given Hamiltonian or energy function. These processors are capable of analog quantum simulation of physical systems [433, 434], even with applications to biology [435, 436], traffic flow [437], machine learning [438, 439], and so on. Digital quantum simulation has also been realized in superconductor-based quantum computers developed by other companies like Google [430], IBM [428, 429, 440-443], Rigetti [441, 442, 444, 445], etc. These superconductor-based devices are cooled in dilution refrigerators in order to be properly operated. QS has also been demonstrated in trapped ion systems developed by IonQ [421, 422].

| Physical systems | Digital | Analog | Hybrid |
|---|---|---|---|
| trapped ion | 4 [421] | 350 [389] | |
| trapped atom | | 60000 [397] | |
| linear optical system | 6 [415] | 4 [380] | |
| liquid NMR | 3 [412] | 3 [379] | |
| nitrogen-vacancy center | 1 [423] | 1 [406] | |
| quantum dot | | 4 [405] | |
| superconductor | 12 [430] | 1800 [434] | 2 [369] |

**Table 2:** Examples of quantum simulators. Systems working at room temperature are marked with red color, while blue color indicates that the corresponding systems are operated at low temperatures. The number of elementary units (ions, atoms, photons, nuclear spins, nitrogen-vacancy centers, quantum dots, qubits in superconducting circuits) actually used for experimental simulation of quantum systems in the corresponding references is provided in the table.

**4 discussions**

Although progress has been made in QC and QS at room temperature, there are significant efforts devoted to quantum dots, superconducting QC, and topological QC, which require ultra-low temperature environments for operation and investigation. Furthermore, a lower temperature is beneficial for better performance of qubits. For example, the qubits based on quantum dot and superconductor both perform better with lower temperatures (Fig. 4). Lastly, some interesting proposals, such as the topological QC based on superfluid $^3$He, require an even lower temperature environment that a dilution refrigerator cannot provide. Therefore, how convenient it would be to realize different low-temperature environments and the corresponding modification to cryostats will be briefly discussed.

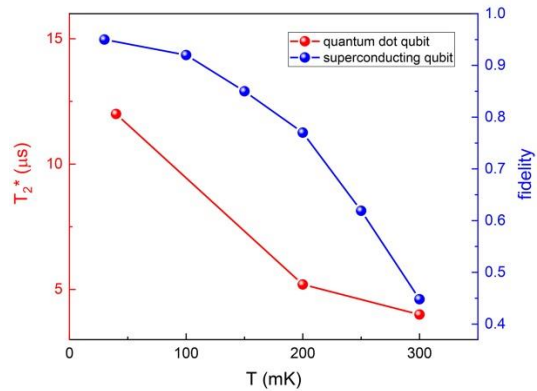

**Figure 4:** Temperature dependence of quantum dot qubit coherence time $T_2^*$ (data from Fig. 4 in Ref. [133]) and superconducting single-qubit gate fidelity (data from Fig. 3c with ν = 245 MHz in Ref. [43]). The temperature in x-axis is the mixing chamber temperature of dilution refrigerators.

Being able to modify existing cryostats helps researchers follow developing technologies and make new breakthroughs. It is relatively easy to realize a low-temperature environment above 2 K with cryocoolers or liquid $^4$He in a lab. An ultra-low temperature environment can be achieved by vaporizing liquid $^3$He, where special care of heat leak is getting more and more important with lower temperatures. A cryostat based on adiabatic demagnetization of paramagnetic salts is an alternative, with additional requirements of extra solenoid and heat switch, but in principle it can only provide non-continuous cooling. The appropriate solution for a 10 mK environment is the dilution refrigerator, which can still be home-made in a lab with low temperature expertise. At 10 mK, the thermometry is not as straightforward as measuring the resistance of a thermometer, and the electron temperature in transport measurements is commonly significantly higher than the refrigerator's base temperature. $^3$He refrigerators, adiabatic demagnetization refrigerators based on paramagnetic salts, and dilution refrigerators are all commercially available, and constructing them in an individual research lab is usually unnecessary. Currently, the ultimate solution for an ultra-low temperature environment is the adiabatic nuclear demagnetization refrigerator pre-cooled by a dilution refrigerator. Adiabatic nuclear demagnetization refrigerators are often constructed by experienced researchers themselves, and < 1 mK environments are available only in a small number of labs.

If the era of extensive application of superconducting QC arrives, at least thousands of qubits may be needed, substantially larger than the number of 53 in Sycamore. In addition, the realization of truly large-scale fault-tolerant quantum computers also requires quantum error correction with logical qubits. Therefore, for a moderate number of qubits, dilution refrigerators with larger cooling power and more leads are needed. A standard dilution refrigerator can hold more than 200 coaxial cables if planned carefully. For thousands of qubits, then new communication techniques between low temperature and room temperature are needed. With the blooming fundamental research at ultra-low temperatures, "tools" in addition to conventional transport measurements, such as filtering, noise measurements, microwave attenuators, strain and stress, tilted magnetic field, thermal-transport, on-chip cooling, may turn out to be attractive occasionally.

# 5 perspectives

The low temperature environment is an important ingredient of QC and QS, from the aspect of realizing significant amounts of types of qubit. Cryocoolers based on electricity can replace the consumption of liquid $^4$He as the thermal bath of ~ 4 K precooling, benefiting users without the training in cryogenics. Nowadays, even the lowest temperature environment from adiabatic nuclear demagnetization can be achieved based on pre-cooling of cryocoolers [446-449].

Part of the fundamental research of QC and current efforts of commercial companies heavily depend on dilution refrigerators. However, dilution refrigerators are mainly provided by a handful of companies. In addition, inappropriate operations could cause the leakage of $^3$He from dilution refrigerators, and $^3$He is expensive and difficult to purchase. Before the quickly increasing purchase of dilution refrigerators with the progress of QC, they generally serve fundamental research with limited demands. If the extensive application of QC in need of ultra-low temperature environment arrives, mass production of attention-free dilution refrigerators is necessary.

Although the dilution refrigerator is the best solution for an ultra-low temperature environment so far, the stable supply of $^3$He is critical for its mass production. Whether the storage and supply of $^3$He is enough to support the new industry of QC is an unknown question or even a question with a negative answer. Therefore, new attentions for continuous and steady ultra-low temperature environments without the requirement of $^3$He are highly desirable.

**Acknowledgements**
We thank Xiongjun Liu, Biao Wu, Jiaojie Yan, and Xinyu Wu for discussions. This work was supported by the National Key Research and Development Program of China (2017YFA0303301), the NSFC (11674009 and 11921005), the Beijing Natural Science Foundation (JQ18002) and the Strategic Priority Research Program of Chinese Academy of Sciences (Grant No. XDB28000000).